# Mass and luminosity evolution of young stellar objects

**Philip C. Myers**

Harvard-Smithsonian Center for Astrophysics, 60 Garden Street, Cambridge

MA 02138 USA

pmyers@cfa.harvard.edu

**Abstract.** A model of protostar mass and luminosity evolution in clusters gives new estimates of cluster age, protostar birthrate, accretion rate and mean accretion time. The model assumes constant protostar birthrate, core-clump accretion, and equally likely accretion stopping. Its parameters are set to reproduce the initial mass function, and to match protostar luminosity distributions in nearby star-forming regions. It obtains cluster ages and birthrates from the observed numbers of protostars and pre-main sequence (PMS) stars, and from the modal value of the protostar luminosity. In 31 embedded clusters and complexes the global cluster age is 1-3 Myr, matching available estimates based on optical spectroscopy and evolutionary tracks. This method of age estimation is simpler than optical spectroscopy, and is more useful for young embedded clusters where optical spectrocopy is not possible. In the youngest clusters, the protostar fraction decreases outward from the densest gas, indicating that the local star-forming age increases outward from a few 0.1 Myr in small protostar-dominated zones to a few Myr in large PMS-dominated zones.

1. **Introduction**

Most stars are born in clusters, spatially concentrated on scales of order 1 pc, over time periods of a few Myr (Lada 2010). The mass distribution of such cluster stars approximates the initial mass function (IMF) derived from field stars (Bastian et al. 2010). However it is still unclear how nature produces clusters with these and other observed properties.

Many descriptions of clustered star formation have been presented, based on analytic models (Fletcher & Stahler 1994a, b, McKee & Offner 2010, Offner & McKee 2011, Myers 2010, 2011), and on numerical simulations (Bate 2009, 2011, Banerjee et al. 2009, Offner et al. 2009, Smith et al. 2009, Wang et al. 2010). Forming clusters may be structured by turbulent flows, self-gravity, and magnetic forces into filamentary regions of high surface density with multiple gravitating centers. Recent reviews of models are given by Pudritz (2010) and Clarke (2010). A quantitative understanding of the distributions of stellar masses and ages in clusters requires understanding how often cluster stars are born, how much mass they gain over time, and when they stop gaining their mass.

This paper addresses the question "what are the star-forming ages of young clusters?" It predicts cluster properties based on a model described earlier (Myers 2011; Paper 1). In this model, protostars have constant birthrate, they grow by core-clump accretion, and their accretion is equally likely to stop at any moment, as explained below. This paper extends the model to PMS stars and to times after cluster formation. It sets parameter values by matching to the initial mass function and to the protostar luminosity function in nearby star-forming regions.

The main result of the paper is a new way to estimate the star-forming age of a cluster from its protostar fraction (PF). This method is simpler than using optical spectroscopy (OS) and evolutionary tracks, and it applies to younger clusters. Large well-studied clusters have PF ages 1-3 Myr, matching OS estimates. In younger clusters where OS estimates are not possible, the PF decreases outward from the densest regions, indicating a progression of star-forming age from a few 0.1 Myr to a few Myr. This property may constrain models of cluster evolution.

## 1.1. Protostar birthrates

This paper assumes that clustered star formation is distributed over a time span of a few Myr, based on observations of young clusters (Palla & Stahler 2000). Optically revealed young clusters have ages and age spreads of at least 1 Myr in their pre-main sequence (PMS) stars, derived from their luminosities, spectral types, and evolutionary tracks on the color-magnitude diagram (da Rio et al 2009, Reggiani et al. 2011). Some of this age spread is due to observational and model uncertainties (Baraffe et al. 2010). However, embedded clusters harbor PMS stars with age ~ 1 Myr and protostars with age ~ 0.1 Myr, as in IC348 (Muench et al. 2007) and in other nearby regions of star formation (Gutermuth et al. 2008, Kryukova et al. 2012). The presence of young stellar objects (YSOs) with such widely varying ages supports the idea that protostar birth in clusters ranges over a few Myr (Fletcher & Stahler 1994a, b; hereafter FS94; Tan et al. 2006). This range probably does not exceed ~ 5 Myr, since most clusters older than 5 Myr have very little associated molecular gas (Leisawitz et al. 1989).

The protostar birth history is assumed to have a sudden start, a constant birthrate, and a sudden stop, as was also assumed by FS94. This description is idealized, but its simplicity allows useful estimates of cluster age and birthrate, as described in Section 5. In future studies it will be more realistic to model the typical birth history as gradually accelerating with a time scale of ~ 1 Myr, due to clump contraction, followed by a more sudden deceleration due to dispersal of star-forming gas by recently formed stars (Palla & Stahler 2000, Huff & Stahler 2006, 2007).

## 1.2. Protostar accretion

It is assumed here that the protostar mass accretion rate has a constant "core" component and a mass-dependent "clump" component. Protostars are frequently associated with dense cores, indicating that cores are birth sites of protostars (Beichman et al. 1986, Enoch et al. 2006, Jørgensen et al. 2008). The collapse of an isothermal dense core leads to a nearly constant mass accretion rate, which can account for the masses and likely formation times of low-mass stars (Shu 1977). However, such an accretion rate is too low to match the masses of more massive stars (Myers & Fuller 1992, Bonnell et al. 1997, McKee & Tan 2003). Additional accretion may be provided by the lower-density filamentary "clump" gas which surrounds cores in star-forming regions. Clump gas is well-studied in maps in spectral lines and in the dust continuum (Ridge et al. 2003, Bergin & Tafalla 2007). This gas appears available for accretion, since core column density profiles merge smoothly into their surrounding gas (Teixeira et al. 2006, Kirk et al. 2006). Furthermore, maps of spectral line asymmetry indicate that clump gas is contracting onto the densest starless cores. Such contracting motions, with speeds $\sim 0.1$ km s$^{-1}$, are more prevalent than motions of expansion or oscillation (Lee & Myers 2011). The contribution of both core gas and clump gas to protostar mass is referred to here as "core-clump accretion."

In this model accretion stops suddenly with no "tapering" of the accretion rate. As with the model of constant protostar birthrate, this simplification is useful because it allows a well-defined, parameter-sparing transition from protostars, which are still accreting, to PMS stars, which have stopped accreting. A tapered decrease of the accretion rate with time has been modeled as an exponential (Myers et al. 1998), as a power law (Smith et al. 1998), and as linear (McKee & Offner 2010, OM11). Comparison of tapered and untapered accretion models indicates that tapering yields slightly increased mean luminosity and mean star formation time (OM11).

### 1.3. Accretion durations

The durations of star-forming accretion are assumed to have a broad distribution because in dense young clusters, the processes which limit accretion act over a broad range of times. The close spacing of protostars and cores suggests that accretion times may vary due to dynamical ejection of protostars, to gas dispersal due to stellar feedback, and to gravitational competition with neighboring accretors (Paper 1).

Recent simulations of young clusters support the idea that protostar accretion times have broad distributions. Smooth particle hydrodynamic (SPH) simulations of cluster formation by turbulent stirring of an initial condensation indicate accretion durations spanning two orders of magnitude, whether radiative feedback heating is neglected (Bate 2009) or included (Bate 2011).

The mass distributions in these simulations differ because heating inhibits fragmentation, but in each case the broad range of masses is correlated with a broad range of accretion times. An adaptive mesh refinement (AMR) simulation with turbulent stirring of a denser initial condensation also gives a duration range of more than an order of magnitude, with or without radiative feedback (Krumholz et al. 2011, Figure 12).

The broad range of accretion durations is due primarily to dynamical ejections in the simulations of Bate (2011), which include radiative heating of the gas but do not include effects of magnetic fields and protostellar outflows. Including outflows enhances effects of radiative feedback, provided the initial surface density is low enough (Cunningham et al. 2011). Including effects of both magnetic fields and outflows again inhibits accretion of massive stars but still provides a broad range of stellar masses (Wang et al. 2010, Li et al. 2010), and provides a level of star formation efficiency which matches cluster observations (Nakamura & Li 2011).

The broad ranges of accretion duration in simulations are well described by the equally likely stopping (ELS) model, where the likelihood of stopping at any moment is constant and independent of time (Basu & Jones 2004, Bate & Bonnell 2005, Myers 2009). In contrast, if these ranges of duration were due to free-fall collapse of isolated cores, their initial mean densities would span up to four orders of magnitude, contrary to observed core properties.

### 1.4. Overview

This model is similar to the cluster model of FS94 in its constant birthrate and probabilistic termination of accretion. However the accretion rate in this model has both a constant and a mass-dependent component, and its accretion luminosity is derived from a constant protostar radius, whereas FS94 have a constant accretion rate and a mass-dependent protostar radius. In relation to the accretion models of Offner & McKee (2011; OM11) the present model has an untapered two-component accretion law which is similar to the two-component turbulent core (2CTC) and two-component competitive accretion (2CCA) models. The constant component of the accretion rate is similar to that of the isothermal sphere (IS) model. The broad distribution of accretion durations in this model differs from the competitive accretion (CA) model of OM11, where YSOs of all masses have the same accretion duration.

This paper develops these ideas of constant birthrate, core-clump accretion, and equally likely stopping to predict properties of protostar and PMS populations in young clusters. It goes beyond Paper 1 and other recent treatments, because it predicts PMS populations in addition to protostar populations, and because it predicts cluster properties after star births have stopped. Section 2 gives distributions of accretion duration, and numbers of protostars and PMS stars, as a function of time. Section 3 gives similar distributions for protostar and PMS masses, and Section 4 gives distributions for protostar accretion luminosity. Section 5 applies these results to nearby young clusters, to estimate their ages and birthrates, and to describe the spatial progression of star-forming age in the youngest clusters. Section 6 discusses the results, and Section 7 concludes the paper.

Readers interested primarily in the application of the model to estimating cluster properties may wish to skip the mathematical development in Sections 2-4.

## 2. Distributions of YSOs and YSO ages

This model has three important times. The accretion age $a$ is the accretion duration, or the time that a YSO has spent accreting since its birth. The cluster age $t$ is the time elapsed since the birth of the first YSO in a cluster, and the star-forming lifetime $t_c$ is the cluster age when its last YSO is born. When $t < t_c$ the cluster is still forming YSOs; when $t > t_c$ the cluster has stopped forming YSOs. For a given accretion model, the accretion age $a$ sets the mass of a YSO. At time $t$ the YSOs in a cluster have maximum possible accretion age $a_{max} = t$, if the first-born YSO is still accreting.

The number of YSOs at $t$ is denoted $N$, and the number of YSOs per interval of accretion age between $a$ and $a+ da$ is denoted $dN/da$. The probability density $p(a)$ that a YSO accretes until $a$ and then stops accreting between $a$ and $a +da$ is related to $dN/da$ by

$$\frac{dN}{da} = Np(a) \quad . \tag{1}$$

As in Paper 1, the probability density *p(a)* is given by

$$p(a) = f \frac{\exp(-a/\bar{a})}{\bar{a}} \qquad (2)$$

where the accretion ages span $0 \leq a \leq a_{max}$ and where $\bar{a}$ is the accretion time scale. Integrating equation (2) sets the normalizing coefficient to $f = [1-\exp(-a_{max}/\bar{a})]^{-1}$. As the maximum accretion age $a_{max}$ becomes much larger than the accretion time scale $\bar{a}$, $f$ approaches unity, the accretion time scale approaches the mean accretion age, and *p(a)* approaches the probability density for $a_{max} \rightarrow \infty$. This limiting case is called equally likely stopping because then the probability density of stopping between *a* and *a + da* is equal to $1/\bar{a}$, independent of *a*.

The normalizing coefficient *f* and the accretion time scale $\bar{a}$ are approximately equal to their limiting ELS values provided $a_{max}$ exceeds $\bar{a}$ by at least a factor of a few. Thus *f* exceeds its limit of unity by less than 10% when $a_{max} > 3\bar{a}$, and $\bar{a}$ exceeds the mean accretion age by less than 10% when $a_{max} > 4\bar{a}$.

This section derives expressions for *N* as a function of *t* and *dN/da* as a function of *a* and *t*. These expressions extend those in Paper 1 for protostars during cluster formation, to protostars and to PMS stars, during and after cluster formation.

### 2.1. Protostars during cluster formation

For constant birthrate, the number of protostars $N_{PS}$ as a function of time during the period of cluster formation is obtained by equating the net rate of gain in the protostar population to the gain rate due to births minus the loss rate due to accretion stopping:

$$\frac{dN_{PS}}{dt} = b - \frac{N_{PS}}{\bar{a}} \quad . \tag{3}$$

Integration of equation (3) gives

$$N_{PS} = b\bar{a}[1 - \exp(-t/\bar{a})] \quad , \tag{4}$$

for $0 \leq t \leq t_c$. Equation (4) shows that the number of protostars increases linearly with time and then approaches the constant value $b\bar{a}$ as the rate of accretion stopping approaches the birth rate. This approach to a steady-state population of protostars is similar to that found by FS94.

At time $t$ since the first protostar birth, the maximum possible accretion duration is $a_{max} = t$, if the first-born protostar is still accreting. Then equations (2) and (4) give the number distribution of accretion ages as

$$\frac{dN_{PS}}{da} = b\exp(-a/\bar{a}) \tag{5}$$

where $0 \leq a \leq t$ and $0 \leq t \leq t_c$. Integration of equation (5) over all accretion ages $0 \leq a \leq t$ reproduces the expression for the number of protostars as a function of $t$ in equation (4).

**2.2 YSOs during and after cluster formation**

The calculations of Section 2.1 for protostars can be extended to PMS stars during cluster formation, since in this model each protostar becomes a PMS star as soon as it stops accreting. Thus summing the distributions of protostar accretion duration in equation (5) over all stopping times between 0 and $t$ gives the distribution of stopped accretion durations for all PMS stars present at time $t$. Integrating this distribution of durations over $0 \leq a \leq t$ gives the number of PMS stars at $t$. The result of this calculation is summarized in Table 1, cases *(1) - (3)*, and in equations (6)-(8).

The foregoing calculations for protostars and PMS stars during cluster formation were extended to the time period after cluster formation, when $t > t_c$ and $b = 0$, when the surviving protostars complete their accretion and their accretion ages approach their final values. These results are also summarized in Table 1, cases *(4) - (6)* and in equations (9)-(10).

These distributions of duration are written in the convenient form

$$\frac{dN}{da} = gb\exp(-a/\bar{a}) \qquad (6)$$

where the coefficient $g$ is given in Table 1 for its corresponding range of time $t$ and duration $a$.

**Table 1**. Distributions of accretion duration

| case | YSO | $t_{min}$ | $t_{max}$ | $a_{min}$ | $a_{max}$ | $g$ |
|------|-----|-----------|-----------|-----------|-----------|-----|
| 1 | PS | 0 | $t_c$ | 0 | $t$ | 1 |
| 2 |    | $t_c$ | $\infty$ | 0 | $t-t_c$ | 0 |
| 3 |    | $t_c$ | $\infty$ | $t-t_c$ | $t$ | 1 |
| 4 | PMS | 0 | $t_c$ | 0 | $t$ | $(t-a)/\bar{a}$ |
| 5 |    | $t_c$ | $\infty$ | 0 | $t-t_c$ | $t_c/\bar{a}$ |
| 6 |    | $t_c$ | $\infty$ | $t-t_c$ | $t$ | $(t-a)/\bar{a}$ |

In Table 1, the times $t_{min}$ and $t_{max}$ specify the time range during cluster formation, when $0 \leq t \leq t_c$, or after cluster formation, when $t_c \leq t < \infty$. At a given time $t$, the dependence of $dN/da$ on $a$ differs according to whether the YSO is a protostar or a PMS star, and whether the time $t$ occurs during or after cluster formation. During cluster formation, protostars and PMS stars have the same range of durations, from $a_{min} = 0$ to $a_{max} = t$, but different expressions for $dN/da$. After cluster formation, the expression for $dN/da$ differs according to whether the YSO is a protostar or PMS star, and also according to whether $a$ is in the range from 0 to $t-t_c$ or from $t-t_c$ to $t$. Note that after cluster formation ($t > t_c$) there can be no protostars with accretion ages less than $t-t_c$ (case 2, $g = 0$), because no protostars were born more recently than the lookback time $t-t_c$.

Equation (6) and Table 1 indicate that in a forming cluster, PMS stars are more numerous than protostars, but protostars have slightly longer accretion ages. These properties are shown in Figure 1 for a 0.5 Myr old cluster with birthrate $b = 300$ YSOs Myr$^{-1}$ and accretion time scale $\bar{a} = 0.1$ Myr. The birthrate and age are chosen to represent a young embedded cluster like Serpens South (Gutermuth et al 2008), and the timescale is chosen to match a recent estimate of the Class 0 + Class I lifetime (Dunham & Vorobyov 2012). More detailed estimates for observed clusters and complexes are given in Section 5.

In Figure 1 the 120 PMS stars have modal accretion age 0.081 Myr while the 30 protostars have modal accretion age 0.10 Myr. At the oldest accretion ages, the protostars also have a shallower decline with increasing age than do the PMS stars.

Table 1 shows that the distributions of accretion age have nearly identical shape for protostars approaching steady state during cluster formation, case *(1)* when $\bar{a} \ll t < t_c$, and for PMS stars long after cluster formation, case *(5)* when $t \gg t_c$. Then the distribution amplitudes differ by the constant factor $t_c/\bar{a}$ and the distribution shapes differ only in the extents of their long-duration tails. This similarity indicates that the steady-state distribution of protostar accretion ages is a useful estimator of the final distribution of YSO accretion ages.

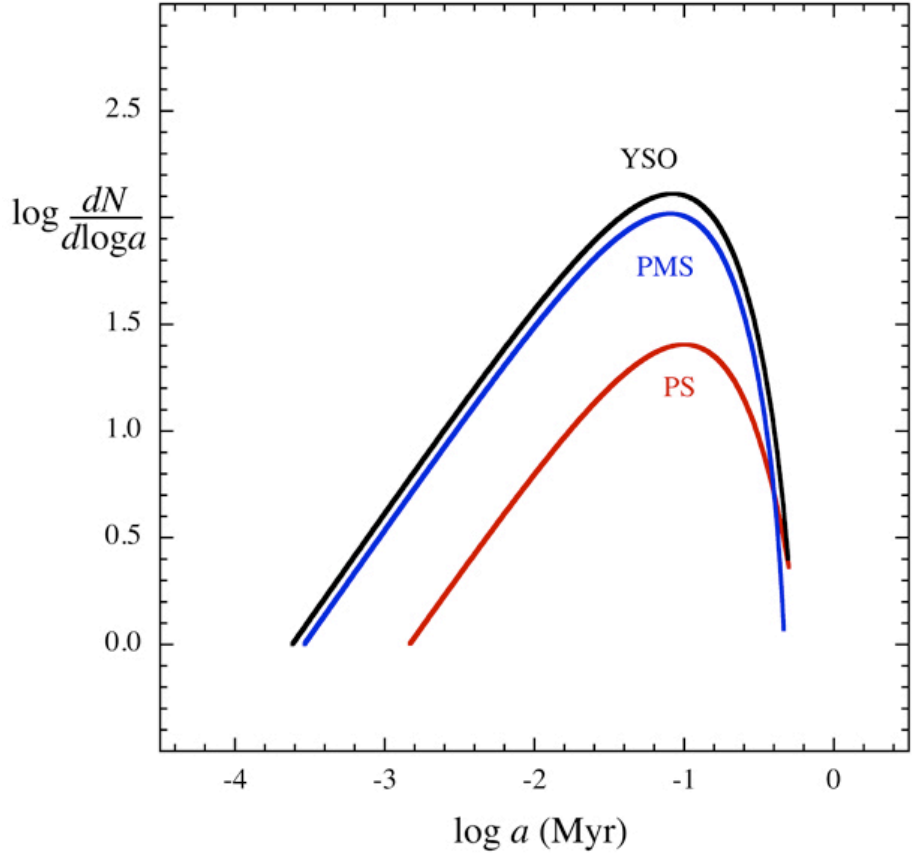

**Figure 1.** Distributions of accretion age $a$ for a forming cluster with birthrate 300 protostars Myr$^{-1}$ and accretion age time scale 0.1 Myr, at cluster age 0.5 Myr since the first protostar was born. The distribution for YSOs *(black)* is the sum of the distributions for protostars, which are accreting *(red)*, and for PMS stars, which have stopped accreting *(blue)*.

The PS and PMS populations $N_{PS}$ and $N_{PMS}$ during cluster formation ($t < t_c$) and after cluster formation ($t > t_c$) are obtained as a function of time $t$ by integrating the distribution of durations in equation (6), over the range of durations in Table 1, giving

$$N_{PS}(t < t_c) = b\bar{a}[1 - \exp(-t/\bar{a})] \qquad (7)$$

$$N_{PMS}(t < t_c) = bt - b\bar{a}[1 - \exp(-t/\bar{a})] \qquad (8)$$

$$N_{PS}(t > t_c) = b\bar{a}[1 - \exp(-t_c/\bar{a})]\exp[-(t - t_c)/\bar{a}] \tag{9}$$

$$N_{PMS}(t > t_c) = bt_c - b\bar{a}[1 - \exp(-t_c/\bar{a})]\exp[-(t - t_c)/\bar{a}]. \tag{10}$$

Equation (7) is identical with equation (4), showing that their derivations are consistent. The number of YSOs is defined to be

$$N_{YSO} = N_{PS} + N_{PMS}, \tag{11}$$

whence the YSO population during cluster formation due to summing equations (7) and (8) is that expected for constant birthrate $b$ during $0 \leq t \leq t_c$,

$$N_{YSO}(t \leq t_c) = bt. \tag{12}$$

Similarly, the YSO population after cluster formation due to summing equations (9) and (10) is that expected for constant birthrate $b > 0$ during $0 \leq t \leq t_c$ and $b = 0$ thereafter,

$$N_{YSO}(t \geq t_c) = bt_c. \tag{13}$$

In a special case, equations (7) and (8) reduce to the well-known equality between population ratio and time scale ratio, used to estimate YSO evolutionary time scales (Evans et al. 2009 and references therein). This result follows when the cluster age is close to its star-forming duration, $t \approx t_c$, and when the star-forming duration is much greater than the accretion time scale, $t_c \gg \bar{a}$. Then $N_{YSO}/N_{PS} \approx t_c/\bar{a}$. If this scaling relation is used to estimate the accretion time scale when a cluster has not yet completed its star formation, it can overestimate the time scale by a factor $t_c/t$.

The number of protostars and PMS stars during and after cluster formation is illustrated in Figure 2, based on equations (7)-(13), and on assumed values $b = 300$ protostars Myr$^{-1}$, $\bar{a} = 0.1$ Myr, and $t_c = 1$ Myr, over the time range $t = 0$-2 Myr. Figure 2 shows that during cluster formation, the PS population approaches a steady value of $b\bar{a} = 30$ protostars, in contrast to the monotonic increases of the PMS and YSO populations. After cluster formation, the protostar population decreases with time scale 0.1 Myr until all the protostars have become PMS stars.

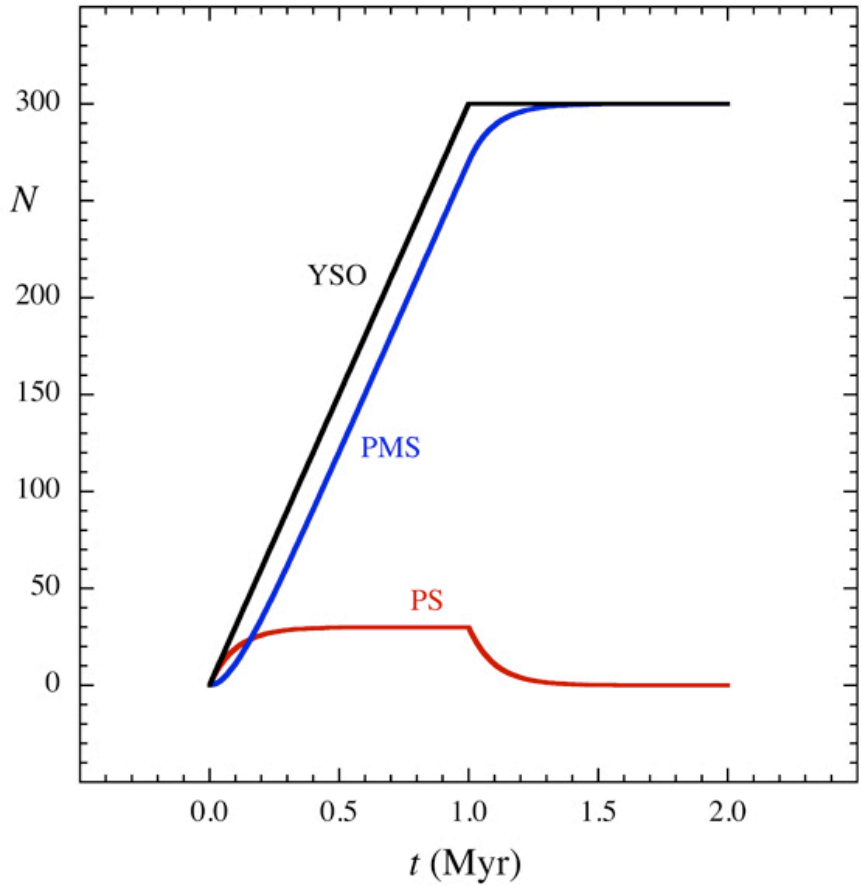

**Figure 2.** The populations $N$ of protostars *(red)*, pre-main sequence stars *(blue),* and young stellar objects *(black)* as functions of time $t$ for a cluster with birthrate $b = 300$ protostars Myr$^{-1}$, over a star-forming lifetime $t_c = 1$ Myr, with accretion time scale $\bar{a} = 0.1$ Myr.

Figure 2 shows that the relative values of the protostar and PMS populations vary with time during cluster formation. The relative protostar and PMS numbers are used to infer ages of embedded clusters in Section 5.

The distributions of accretion duration described here are the basis of the mass and accretion luminosity distributions presented in Sections 3 and 4. When the distributions of accretion duration are written in the logarithmic form $dN/d\log a$ as in Figure 1, their shapes are variants of the function $(a/\bar{a})\exp(-a/\bar{a})$ which has a linear rise, a peak at $a = \bar{a}$, and a steep decline. This function is relatively broad: its half-maximum arguments 2.68 and 0.23 have ratio 11.6. These properties indicate that the broad, single-peak nature of the mass and luminosity functions obtained in Sections 3 and 4 arises from these same features in the distribution of accretion ages.

## 3. Distributions of protostar and PMS masses

Distributions of protostar and PMS star masses follow from the distributions of accretion age given in Section 2, based on relations between accretion age and mass for core-clump accretion (Paper 1). In this accretion model, the mass accretion rate has a constant "core" component $\dot{m}_{core}$ and a mass-dependent "clump" component whose sum is

$$\dot{m} = \dot{m}_{core}(1 + \mu^p) \qquad (14)$$

where the normalized YSO mass is

$$\mu \equiv \frac{m}{m_0}.\qquad(15)$$

Here the mass scale is $m_0 = \dot{m}_{core}\tau_{clump}$ and $\tau_{clump}$ is the time scale for accretion from the clump gas onto the protostar. In equation (14) the mass exponent $p$ takes the value $p = 1.2$ which provides best fit of the final mass function to the IMF. It was found that an accretion rate with two components gives a better fit to the IMF and to observed luminosity functions than either a pure "core" accretion rate or a pure "clump" accretion rate (OM11; Paper 1).

Integration of equation (14) gives the relation

$$\frac{a}{\tau_{clump}} = \int_0^\mu \frac{d\mu'}{1+\mu'^p}.\qquad(16)$$

Here the normalized mass $\mu$ increases with accretion duration, as detailed in Paper 1.

### 3.1. Mass function evolution

The distributions of duration in Section 2 and the relation of protostar mass to duration in equation (16) give distributions of protostar and PMS mass as functions of mass and time, according to

$$\frac{dN}{dm} = \frac{dN}{da}\frac{da}{dm}\qquad(17)$$

for *dN/da* as given in equation (6) and Table 1. In terms of the logarithmic derivative, the resulting mass distributions are written as

$$\frac{dN}{d\log m} = g\bar{b}\bar{a}h_m \quad (18)$$

where

$$h_m \equiv \ln(10)\frac{q\mu}{1+\mu^p}\exp\left(-q\int_0^\mu \frac{d\mu'}{1+\mu'^p}\right) \quad (19)$$

and where $q = \tau_{clump}/\bar{a}$ is the ratio of time scales. In equation (18) the dependence of $g$ on time and mass is given by Table 1 and by equation (16). In equation (19) the local maximum of $h_m$ occurs when $\mu \equiv \mu_{0m} = 0.461$.

Comparison of equations (6) and (18) shows that the shape of the mass function derives its low-mass rise, its peak, and its broad width from the distribution of durations, as discussed in Section 2.2. The high-mass tail of the mass function comes from the clump component of accretion onto the protostar, which causes the exponential term in equation (19) to decline less steeply as $p$ increases. Equation (16) for protostars during cluster formation is the same as equation (19) of Paper 1, using equation (1) of this paper and allowing for small differences in notation.

The mass distribution in equation (18) can be written simply as

$$\frac{dN_{PS}}{d\log m} = N_{PS} h_m \qquad (20)$$

when the protostar population is close to its steady state, when $b\bar{a} = N_{PS}$, and as

$$\frac{dN_{YSO}}{d\log m} = N_{YSO} h_m \qquad (21)$$

for YSOs after cluster formation. Equations (20) and (21) show that the mass distributions for these cases *(1)* and *(5)* in Table 1 have essentially the same shape, and a constant amplitude ratio $N_{YSO}/N_{PS}$. Their high-mass tails differ in extent as given in Table 1.

As a cluster forms, its protostar and PMS mass functions based on equation (18) evolve differently. These mass functions were calculated for a constant birthrate cluster forming $b = 300$ protostars $\mathrm{Myr}^{-1}$ over $t_c = 1$ Myr. Their parameters of core-clump accretion and ELS were chosen to give a protostar steady-state mass function having the same shape as the IMF. These parameters are $m_0 = 0.34\ M_\odot$, $q = 2.0$, and $p = 1.2$ (Paper 1). The resulting mass functions for protostars, PMS stars and YSOs are shown in Figures 3 and 4 for times $t = 0.5$ and 1.0 Myr.

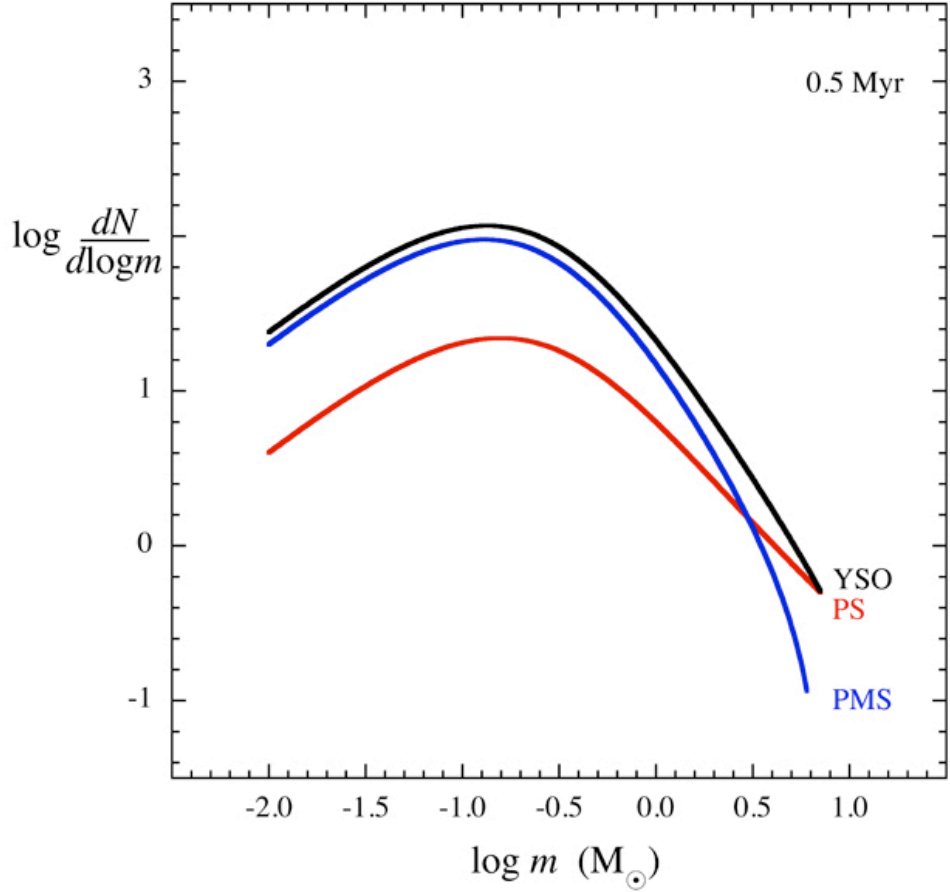

**Figure 3.** Mass distributions of protostars *(red)*, pre-main sequence stars *(blue)*, and young stellar objects *(black)* in a forming cluster with birthrate $b = 300$ stars $Myr^{-1}$ and accretion timescale $\bar{a} = 0.1$ Myr, at cluster age $t = 0.5$ Myr. The core-clump accretion parameters are $m_0 = 0.34\ M_\odot$, $q = 2.0$, and $p = 1.2$.

Figures 3 and 4 show that at cluster age 0.5 Myr, PMS stars have a slightly smaller modal mass, a greater modal number, and a steeper decrease of their number with increasing mass than do protostars. These features are due to the ELS property that short accretion durations are more likely than long durations, and to the monotonic increase of mass with duration. The relationship of the PS and PMS modal masses, and the relationship of their modal numbers, is essentially the same at 1.0 Myr as at 0.5 Myr. The shape and amplitude of the PS mass function are nearly the same at 1.0 Myr as at 0.5 Myr, since at these times the PS distribution is close to its steady state.

On the other hand PMS stars are more numerous, and their high-mass tail is higher at 1.0 Myr than at 0.5 Myr, since more stars of all masses have formed, and since more massive protostars have stopped accreting. The shapes of the PMS and PS mass functions are more similar at 1.0 Myr than at 0.5 Myr. Their amplitude ratio has increased from 4.6 to 9.3, approaching the value $t_c / \bar{a} = 10$ expected for times long after cluster formation.

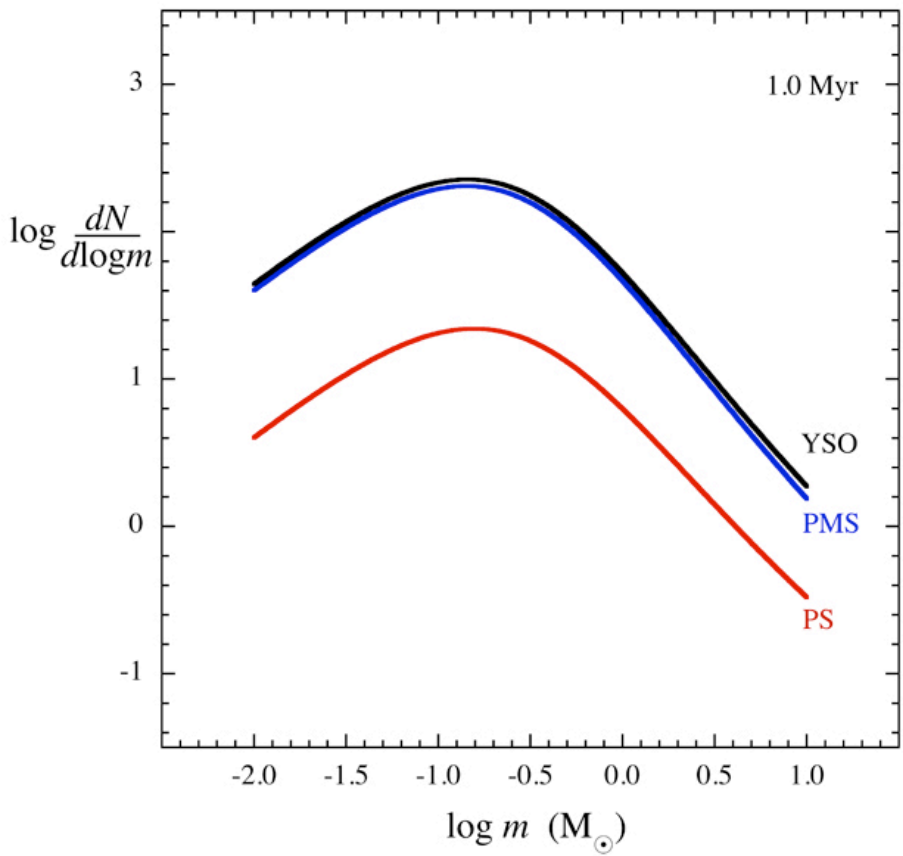

**Figure 4.** Mass distributions of protostars *(red)*, pre-main sequence stars *(blue)*, and young stellar objects *(black)* in a forming cluster with birthrate $b = 300$ stars Myr$^{-1}$ and accretion

timescale $\bar{a} = 0.1$ Myr, at cluster age $t = 1.0$ Myr. The core-clump accretion parameters are $m_0$ = 0.34 $M_\odot$, $q = 2.0$, and $p = 1.2$.

At cluster age 1.5 Myr, it is expected that no protostars are still accreting, since the last protostar was born 0.5 Myr ago, or five accretion times scales ago. Thus in equation (9) $N_{PS}(t>t_c) << 1$. All of the YSOs are now PMS stars, and the YSO mass function is a final mass function. Figure 5 compares the YSO mass functions at 1.5 Myr with those from 0.5 and 1.0 Myr, showing that the modal masses have become slightly greater and that the high-mass tail has become higher. Each of these changes is due to the conversion of more massive protostars into PMS stars.

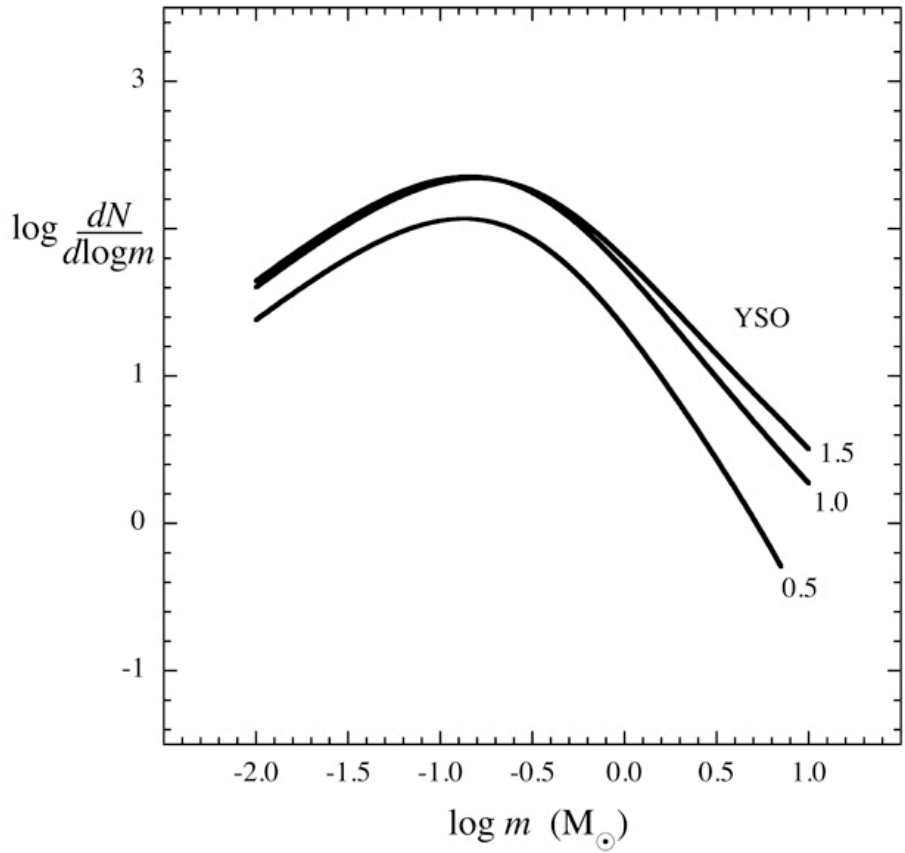

**Figure 5.** Evolution of mass distributions of YSOs in a cluster whose birthrate is $b = 300$ protostars $\text{Myr}^{-1}$ over time period $t_c = 1$ Myr, and whose accretion timescale is $\bar{a} = 0.1$ Myr. The core-clump accretion parameters are $m_0 = 0.34\ M_\odot$, $q = 2.0$, and $p = 1.2$. The mass distributions evolve toward increasing numbers, and toward an increasing proportion of massive stars, from times $t = 0.5$ to $1.0$ to $1.5$ Myr after the first protostar birth.

This final YSO mass function resembles the PS mass function during cluster formation, as shown in Figure 6, and as expected from the discussion in Section 2.2. Figure 6 also verifies that each of these agrees well with the IMF, as expected from the choice of parameters $m_0$, $q$ and $p$.

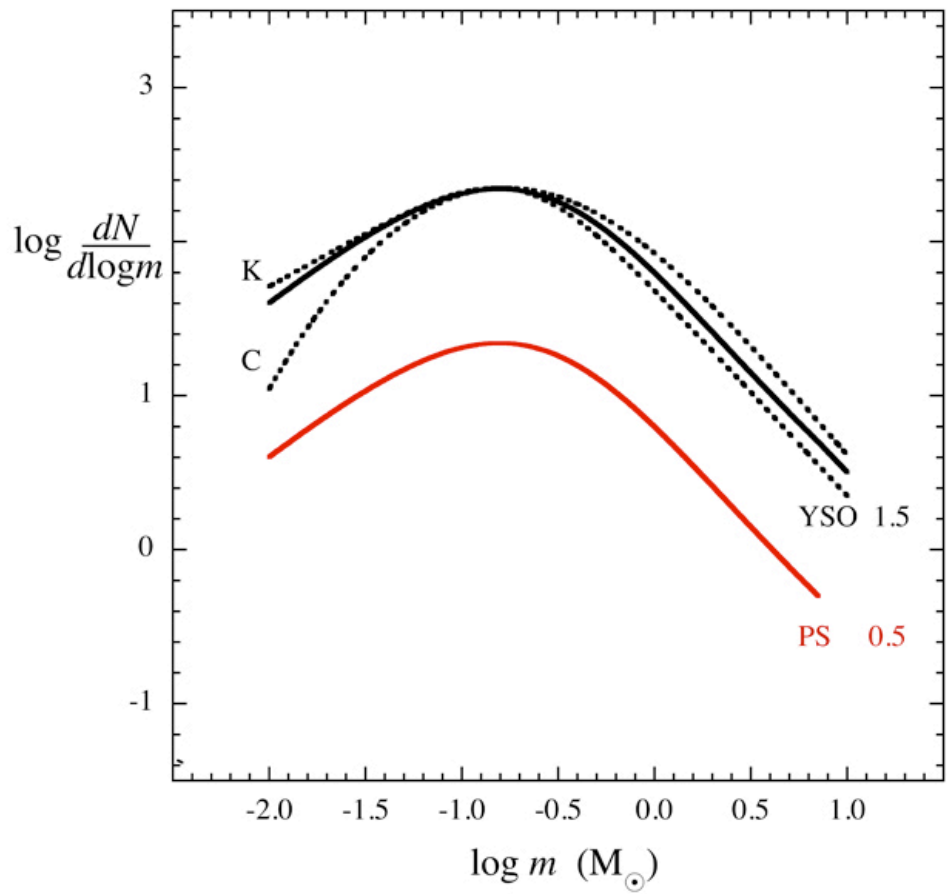

**Figure 6.** Mass distributions of protostars at cluster age 0.5 Myr and YSOs at 1.5 Myr, in a cluster which forms protostars for 1.0 Myr, with birthrate $b = 300$ protostars Myr$^{-1}$ and accretion timescale $\bar{a} = 0.1$ Myr. The core-clump accretion parameters are $m_0 = 0.34\ M_\odot$, $q = 2.0$, and $p = 1.2$. The PS and YSO mass distributions have similar shape. The YSO mass distribution at 1.5 Myr closely matches the IMFs of Kroupa (2002, $K$) and Chabrier (2005, $C$).

### 3.2. Modal mass

The modal value of the mass function is obtained by evaluation of equation (21), giving

$$m_{mod} = \mu_{0m} \dot{m}_{core} q \bar{a} \tag{22}$$

or $m_{mod} = 0.16\ M_\odot$ for the parameters used above. This value is approximated well by the assumption that the mass accretion is due only to its core component $\dot{m}_{core}$. In that case $m_{mod} = \dot{m}_{core}\bar{a} = 0.17\ M_\odot$ for the same parameters. This approximation works well because the peak of the mass function is due to low-mass stars, for which clump accretion is negligible.

### 3.3. Mean and median mass

Figures 3 and 4 show that the modal mass is essentially unchanged as the cluster evolves, once a few accretion time scales have elapsed so that the mass function $dN/d\log m$ has a well-defined peak. In contrast, the maximum mass increases strongly with time, as the high-mass tail of the mass function extends to greater mass values. The mean and median masses also increase with time, but these increases resemble that of the modal mass, and are relatively small once the distribution has a well-defined peak.

This time dependence can be described by simple functions for the special case where core accretion is constant and clump accretion is negligible. Then the mean mass is

$$\overline{m} = \dot{m}_{core}\overline{a}\left[1 - \frac{t/\overline{a}}{\exp(t/\overline{a})-1}\right] \qquad (23)$$

and the median mass is

$$m_{med} = \dot{m}_{core}\overline{a}\ln\left[\frac{2}{1+\exp(-t/\overline{a})}\right] . \qquad (24)$$

These equations indicate that the mean mass lies within 10% of its limiting value $\dot{m}_{core}\overline{a}$ when $t > 3.6\ \overline{a}$, and that the median mass lies within 10% of its limiting value $\dot{m}_{core}\overline{a}(\ln 2)$ when $t > 2.6\ \overline{a}$. When the clump component of accretion is included, the mean and median mass must be computed numerically. However their approach to a constant value after a few accretion time scales is similar to that shown in equations (23) and (24).

### 4. Luminosity function evolution

For protostars the dominant form of luminosity is accretion luminosity, which is written in terms of the dimensionless mass $\mu$ as

$$L = L_0\mu(1+\mu^p) \qquad (25)$$

where the luminosity scale is

$$L_0 \equiv \frac{\gamma G \dot{m}_{core}^2 \tau_{clump}}{R_*} \quad (26)$$

and where $\gamma$ is the accretion luminosity efficiency with respect to perfect spherical accretion, with the effective protostar radius $R_\star = 2.5\ R_\odot$ (Hosokawa et al. 2010). The accretion luminosity efficiency was estimated as $\gamma = 0.5$ by OM11 and in Paper 1, corresponding to a modest degree of episodic disk accretion. In all following instances, $\gamma$ and $R_\star$ occur in the same combination $\gamma/R_\star$, so it is useful to define the parameter

$$\alpha_L \equiv \frac{\gamma G}{R_*} \quad (27)$$

whose value is $\alpha_L = 1.92 \times 10^{-19}$ erg g$^{-2}$ assuming $\gamma = 0.5$. Here $G$ is the gravitational constant. The logarithmic distribution of accretion luminosity is obtained from the logarithmic mass distribution in equation (18) according to

$$\frac{dN_{PS}}{d\log L} = \frac{L}{m}\frac{dm}{dL}\frac{dN_{PS}}{d\log m}, \quad (28)$$

yielding an expression similar to equation (18),

$$\frac{dN_{PS}}{d\log L} = g\bar{b}\bar{a}h_l \quad (29)$$

where

$$h_l = \ln(10)\frac{q\mu}{1+(p+1)\mu^p}\exp\left(-q\int_0^\mu \frac{d\mu'}{1+\mu'^p}\right). \tag{30}$$

When $\mu \gg 1$, this function $h_l$ declines slightly more rapidly with $\mu$ than does $h_m$ in equation (19). Here $g$ and $\mu$ depend on time, luminosity and luminosity range following equations (18) and (22), and following Table 1 for protostars (cases *1-3*).

In equation (28), $h_l$ has modal value $h_{0l} = 0.532$ when $\mu = \mu_{0l} = 0.347$, for the adopted parameter values $p = 1.2$ and $q = 2.0$. Then equations (25) and (27) give the modal luminosity as

$$L_{\text{mod}} = \beta_L \dot{m}_{core}^2 \bar{a} \tag{31}$$

where

$$\beta_L \equiv \frac{\gamma G q \mu_{0l}(1+\mu_{0l}^p)}{R_*}, \tag{32}$$

or where $\beta_L = 1.71 \times 10^{-19}$ erg g$^{-2}$ for adopted parameter values. The modal value of the luminosity distribution is then

$$\left(\frac{dN_{PS}}{d\log L}\right)_{\text{mod}} = g\bar{b}\bar{a}h_{0l} \tag{33}$$

In the present model of constant birthrate and equally likely stopping, the distribution of protostars approaches a steady state, where the luminosity distribution has a fixed shape. This fixed shape implies that the peak of the distribution and the total number of protostars $N_{PS}$ are linearly related, since $N_{PS}$ is proportional to the area under the distribution. The fixed shape also implies that the linear relation has the same coefficient for all clusters which are near steady state, regardless of their number of members. Since $g = 1$ for protostars during cluster formation, equations (7) and (33) give this linear relation as

$$\left(\frac{dN_{PS}}{d\log L}\right)_{mod} = h_{0l} N_{PS} \quad (34)$$

where $h_{0l} = 0.532$ as obtained above.

Equation (34) is a model property which should be satisfied for all regions whose luminosity distribution is well fit by equation (29). This applies to protostar luminosities in nearby star-forming clouds (Dunham et al. 2010), as shown in Paper 1, and in Orion A, as shown in Section 5.1. In each of these cases equation (34) applies within 10%. Equation (34) may also provide a useful test of the model for embedded clusters whose number of protostars is large enough to give a well-determined modal value, but not large enough to give a statistically significant number of detections in its higher-luminosity bins.

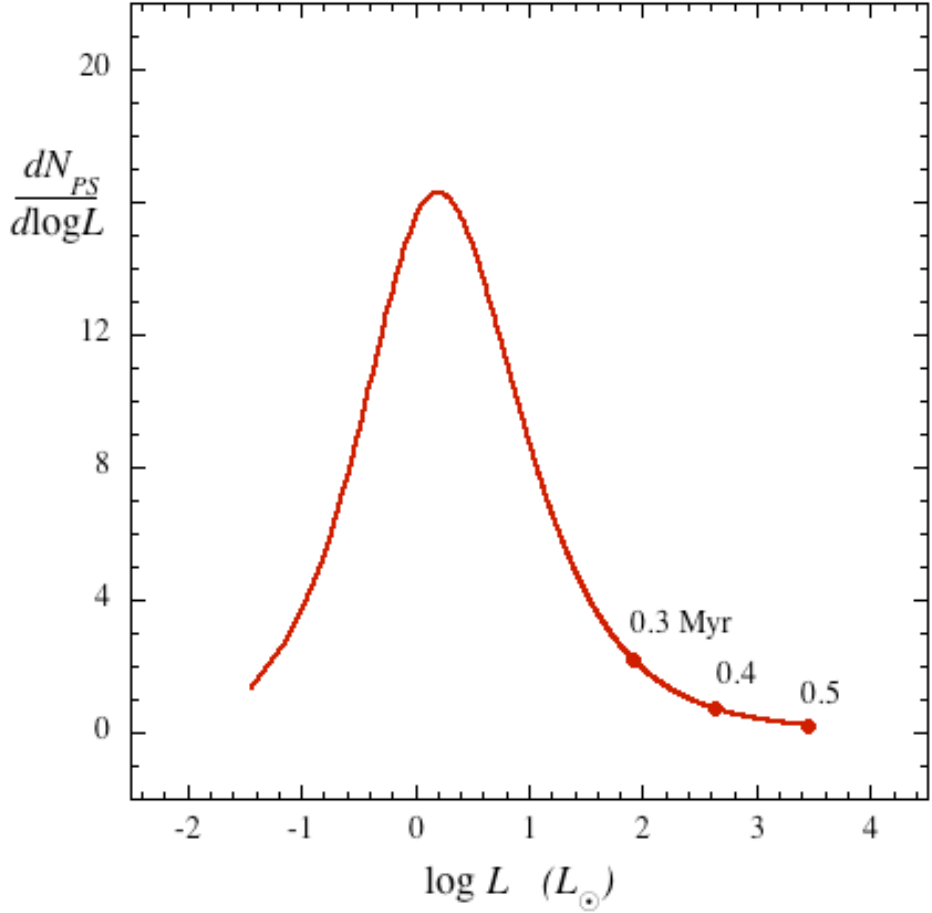

**Figure 7.** Distribution of protostar luminosity in a cluster whose maximum accretion age is 0.3, 0.4, and 0.5 Myr, for birthrate $b = 300$ protostars $Myr^{-1}$ and accretion timescale $\bar{a} = 0.1$ Myr. The core-clump accretion parameters are $m_0 = 0.34\ M_\odot$, $q = 2.0$, and $p = 1.2$. The distribution has a high-luminosity tail, whose maximum luminosity grows with increasing accretion age.

Evolution of the distribution of accretion luminosity based on equation (29) is shown in Figure 7 for the same parameters as in previous examples. Once the cluster age $t$ is at least a few times the accretion time scale $\bar{a}$, the protostar population is nearly in steady state, and the luminosity distribution changes only in the extent of its high-luminosity tail. The modal luminosity is 1.6 $L_\odot$ as given by equation (31), due to the large fraction of low-mass protostars.

The ratio of modal luminosity to modal mass is 10 in solar units, and it depends significantly only on the constant component of the mass accretion rate, $\dot{m}_{core}$.

As the cluster age increases so do the maximum accretion age $a_{max}$ and the maximum accretion luminosity $L_{max}$. Figure 7 shows the luminosity distribution for maximum accretion ages 0.3, 0.4, and 0.5 Myr, when the corresponding maximum luminosities are respectively 81, 430, and 2800 $L_\odot$. For comparison the maximum protostar luminosity in the c2d clouds, in Mon R2, in CepOB3, and in Orion A (excluding the ONC) are respectively 76, 210, 410, and 490 $L_\odot$ (Evans et al. 2009, Kryukova et al. 2012).

## 5. Estimating cluster properties

This section estimates cluster properties by applying the luminosity model of Section 4 to observations of protostars, PMS stars, and protostar luminosities in young clusters.

### 5.1. Fitting protostar luminosity distributions

The distribution of accretion luminosity in equation (29) matches well to the observed distribution of 229 protostar luminosities in Orion A, based on mid-infrared *Spitzer* observations and extrapolation to longer wavelengths (Kryukova et al. 2012). These detailed observations of Orion offer the first opportunity to match a model to the protostar luminosities in one star-forming region. Previous comparisons of models to observations have used luminosities of protostars combined from several nearby clouds at different distances (Offner & McKee 2011, Paper 1). The Orion distribution may be characteristic of complexes forming massive stars, since it resembles the distributions for Cep OB3 and Mon R2 in its asymmetric shape, its high-luminosity tail, and its modal value near 1 $L_\odot$ (Kryukova et al. 2012).

In comparing luminosities from an accretion model with observations, it is important to distinguish luminosity due to accretion from other sources such as PMS contraction or nuclear burning (Dunham et al. 2010). In practice luminosity due to PMS contraction or nuclear burning is less by at least an order of magnitude than accretion luminosity for low-mass YSOs. Further, the identification of the YSO as an accreting protostar as opposed to a more evolved object is well-established on the basis of its spectral energy distribution (SED). The accreting envelope causes a protostar SED to peak in the far infrared, while SEDs of more evolved YSOs peak in the optical or near-infrared (Lada & Wilking 1984, Adams et al. 1987, Myers & Ladd 1993, Evans et al. 2009). If a PMS star is sufficiently massive, approaching $\sim 10\ M_\odot$, its luminosity resembles the accretion luminosity for the same mass (Palla & Stahler 1999). If it also has an edge-on disk to obscure its short-wavelength emission and re-radiate in the far infrared, it can resemble a massive protostar in both luminosity and SED shape. However, such cases are relatively rare.

A model fit to the Orion A luminosities is shown in Figure 8. The model was fit so that at late times its mass distribution also matches the IMF, with the same values of $m_0$, $p$, and $q$ as in Figure 3 of Paper 1. These parameter values fix the shape of the model luminosity distribution. The modal luminosity $L_{mod}$ and amplitude $(dN/d\log L)_{mod}$ were adjusted by eye so that the model curve passes through as many histogram bins as possible. The fit shown matches the observed histogram within $\sqrt{N}$ uncertainty for nearly all histogram bins.

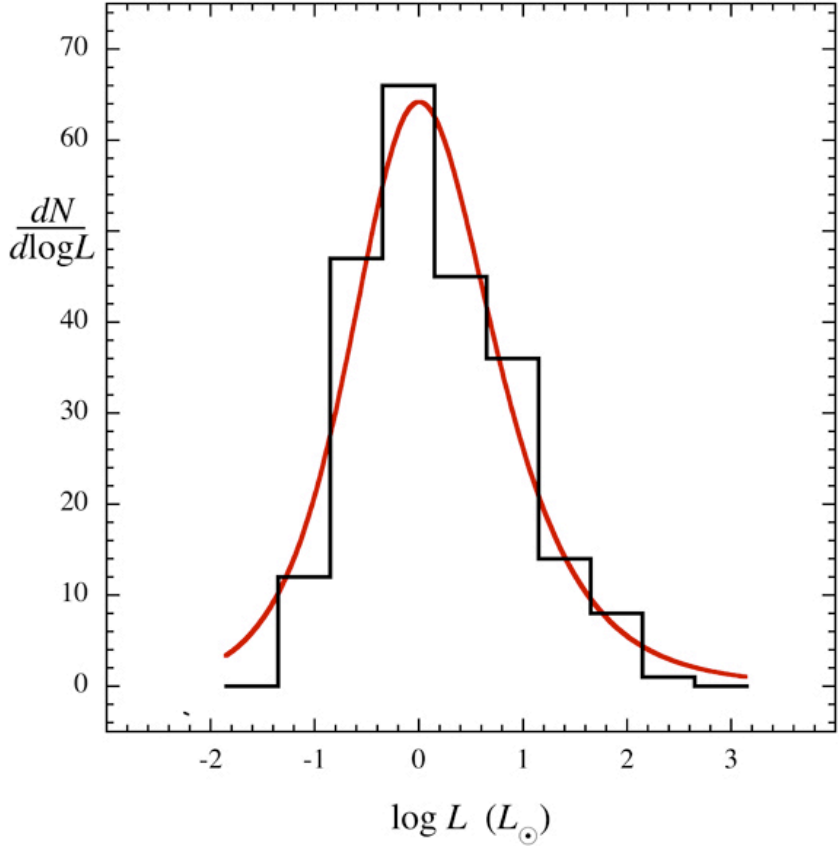

**Figure 8.** Protostar luminosities in the Orion A cloud. *Histogram*, *Spitzer* and 2MASS infrared observations of 229 protostars (Kryukova et al. 2012). *Curve*, model distribution with birthrate $b = 1500$ protostars Myr$^{-1}$, accretion timescale $\bar{a} = 0.16$ Myr, and core-clump accretion parameters $m_0 = 0.34\ M_\odot$, $q = 2.0$, and $p = 1.2$.

The model and its best-fit parameter values indicate that the mean protostar birthrate in Orion A is $b = 1500$ protostars Myr$^{-1}$, the accretion time scale is $\bar{a} = 0.16$ Myr, the core mass accretion rate is $\dot{m}_{core} = 1.1\ M_\odot$ Myr$^{-1}$, and the most massive protostar has luminosity $1500\ L_\odot$. Implications of these properties are discussed below.

The 229 protostars in Orion A are an order of magnitude more numerous than in the typical nearby embedded cluster (Gutermuth et al. 2008). This can be understood since the Orion A complex extends over ~ 50 pc and consists of multiple groups and clusters (Bally 2008). Consequently the protostar birthrate in Orion A, 1500 protostars $Myr^{-1}$, is much greater than the typical birthrate for nearby individual clusters, 50-100 protostars $Myr^{-1}$ (Muench et al. 2007) However the birthrate in Orion A is comparable to the birthrate of the ONC alone, estimated to be ~ 900 protostars $Myr^{-1}$ (Reggiani et al 2011).

The accretion time scale $\bar{a}$ = 0.16 Myr is by definition the typical duration of accretion. Its value inferred here is comparable to time scales for low mass star formation, including the estimated Class 0 + Class I lifetime, 0.12 Myr (Dunham & Vorobyov 2012), the Class 0 lifetime, 0.17 Myr (Enoch et al. 2009) and the estimated star formation time, 0.3 Myr (Offner & McKee 2011). However its value is less than Class 0 + Class I lifetime of 0.4-0.5 Myr, estimated from population ratios in the c2d survey (Evans et al. 2009). As noted in Section 2.2, this use of population ratios may overestimate the accretion timescale if the typical cluster age in the sample is less than the assumed star-forming lifetime.

The core mass accretion rate onto the protostar $\dot{m}_{core}$ = 1.1 $M_\odot$ $Myr^{-1}$ matches the accretion rate found in a detailed study of the low-mass protostar TMC-1 (Terebey et al. 2006). This study takes into account the reduction of the solid angle of infall onto the protostar from its maximum value of 4π, due to dispersal of infalling gas by the protostellar outflow (Velusamy & Langer 1998, Terebey et al. 2006, Myers 2008). This inferred value of $\dot{m}_{core}$ is 30% less than expected for collapse of a singular isothermal sphere at 10 K (Shu 1977).

These inferred values of accretion time scale and mass accretion rate are characteristic of low-mass star formation, since they are set by the the modal mass of the IMF, and by the modal luminosity of the Orion A distribution in Figure 10, which is set primarily by low-mass stars.

The most luminous protostar predicted by the model in Figure 8, whose histogram bin occupation number is unity, has $L$ = 1500 $L_\odot$, while the most luminous observed protostar has $L$ = 490 $L_\odot$ (Kryukova et al. 2012). This discrepancy may be due to the tendency for the most luminous sources to be missed because they appear in regions of extended saturation at 24 $\mu$m wavelength, especially near the ONC (Kryukova et al. 2012). It will be important to compare this model with observations which are not biased against high-luminosity protostars.

The cluster age of the region can be estimated using equation (12) with the estimated birthrate and the total number of YSOs. This number is $N_{YSO}$ = 2300 due to Class I and Class II source identifications from infrared observations (Kryukova et al. 2012). This estimate includes very few sources from the ONC, the most prominent cluster in the Orion A region. Most ONC sources could not be identified due to saturation. This estimate also does not account for Class III sources with undetected infrared excess, identified by x-ray observations. Such x-ray observations of embedded clusters yield a number of Class III sources approximately equal to the number of Class II sources in Cha I (Luhman et al. 2008) and in Cr A (Peterson et al. 2011). It is assumed here that the numbers of Class III and Class II YSOs are equal; this assumption yields a star-forming age estimate of 3.1 Myr for the Orion A region, excluding the ONC.

This age estimate of 3.1 Myr is proportional to the ratio of YSOs to protostars, as discussed in Section 5. Uncertainty in this ratio is likely dominated by incompleteness in detecting and identifying faint and confused PMS stars and protostars. The simple assumption of constant birthrate may also be a source of error if the true birth history is accelerated or episodic. These sources of systematic error probably dominate over random errors, and could underestimate the effective age by a factor of order 2.

This star-forming age of Orion A is similar to the estimated star-forming age of the ONC. The ONC began its star formation about 3 Myr ago, and stopped 1 Myr ago, based on spectroscopic observations and evolutionary tracks (Reggiani et al. 2011).

Apart from those in Orion, relatively few distributions of observed protostar luminosity are available with sufficient completeness for detailed comparison of the luminosity distribution against models (Dunham et al. 2010, Kruyukova et al. 2012), as was done by OM11, in Paper 1, and in this paper. However, more are expected as recent observations with the *Spitzer Space Telescope* and the *Herschel Space Observatory* are analyzed. It should become possible to fit model distributions of protostar luminosity as in Figure 8, to compare from one cluster to the next the estimated cluster age, protostar birth rate, accretion time scale, core mass accretion rate, and the mass and luminosity of the most massive protostar.

## 5.2. Applying modal luminosities and YSO source counts

In many cases the observed distribution of protostar luminosities is not known well enough to allow a significant model fit, but the modal luminosity is known, or it can be plausibly assumed. This assumption is supported by the finding that the regions Orion A, Cep OB3, and Mon R2, each having more than ~ 100 protostars in a recent study of nearby clusters, have luminosity distributions with similar shape, and modal value near 1 $L_\odot$ (Kryukova et al. 2012). This section applies the equations of Section 4 to estimate cluster properties from the modal luminosity $L_{mod}$, the number of protostars $N_{PS}$ and the number of YSOs $N_{YSO}$.

In these estimates, it is assumed as above that the late-time mass distribution of YSOs matches the IMF, with modal mass $m_{mod} = 0.16\,M_\odot$ (Chabrier 2005), that the effective radius of the protostar is 2.5 $R_\odot$ (Hosokawa et al. 2010), and that the efficiency of accretion luminosity with respect to spherical accretion is $\gamma = 0.5$, as expected for a modest degree of episodic disk accretion (OM11). Then equations (25), (34), and (35) give the core component of the mass accretion rate as

$$\dot{m}_{core} = \frac{q\mu_{0m}L_{mod}}{\beta_L m_{mod}} \quad , \qquad (35)$$

and the mean accretion age as

$$\bar{a} = \frac{\beta_L m_{\text{mod}}^2}{(\mu_{0m}q)^2 L_{\text{mod}}} \quad . \tag{36}$$

The cluster age $t$ follows from the protostar fraction $v \equiv N_{PS}/N_{YSO}$ using equations (7) and (12), which give

$$v = \frac{1 - \exp(-t/\bar{a})}{t/\bar{a}} \tag{37}$$

as in equation (20) of Paper 1. Equation (37) does not have an exact solution for $t$, but an approximate solution can be written

$$t \cong \frac{\bar{a}}{v\left[1 + \frac{v^2}{2(1-v)}\right]} \tag{38}$$

with error less than 1% for $0 < t/\bar{a} < 10$. In equation (38) the term $v^2/[2(1-v)]$ becomes negligible as $t/\bar{a}$ increases, or as $v$ decreases: it is less than 0.03 provided $v$ is less than 0.2.

The protostar birthrate then follows from equation (12),

$$b = \frac{N_{YSO}}{t} \quad . \tag{39}$$

The number of protostars and YSOs in a cluster depends on the definition of the cluster outer boundary, which in turn depends on the membership criteria and on observational sensitivity and resolution (Gutermuth et al 2009, Bressert et al 2011). For clusters whose protostars and PMS stars are sufficiently well-mixed, the protostar fraction is relatively independent of spatial extent and thus the age in equation (38) is robustly determined.

On the other hand, in some nearby clusters the protostar fraction is much greater in small dense regions than in their more extended, less dense surroundings. In these clusters it is useful to distinguish the "local" protostar fraction evaluated over a specified subregion from the "global" protostar fraction evaluated over the entire cluster, and to distinguish the corresponding "local" and "global" star-forming ages. Section 5.3 presents estimates of global star-forming ages in a large sample of embedded clusters, and Section 5.4 discusses the progression of local star-forming ages in a few very young clusters.

## 5.3. Ages and birthrates of cluster-forming regions

These equations (38) - (42) are useful to compare properties of star-forming regions for which the PS and PMS populations are known.

A sample of embedded clusters and complexes was selected for comparison with the foregoing model of constant birthrate, core-clump accretion, and equally likely stopping, assuming that the final mass distribution matches the IMF and that the modal protostar luminosity is 1.0 $L_\odot$. For these assumptions the equations of Section 5.2 indicate that the accretion time scale is 0.17 Myr and the core component of the accretion rate is 1.0 $M_\odot$ Myr$^{-1}$. The model equations (40) and (42) are used to determine the global age $t$ and protostar birthrate $b$ for each region, based on the observed number of protostars $N_{PS}$ and the number of Class II YSOs $N_{II}$. These are listed in Table 2. The same equations can also be expressed as a series of curves of constant birthrate and of constant cluster age (isochrones), for comparison with $N_{PS}$ and $N_{II}$. These are shown in Figure 9.

Clusters and complexes were selected for comparison with the model if they were observed in the infrared and mid-infrared with enough detail to detect and discriminate protostars and PMS stars by their assignment into evolutionary classes, as first described by Lada & Wilking (1984) and more recently by Evans et al. (2009). Only regions observed by the Spitzer Space Telescope within a few kpc distance were considered. At least 35 members were required, following the cluster criterion of Lada & Lada (2003). Following usual practice, protostars were considered to have "Class 0," "Class I," "flat-spectrum" or "rising-spectrum" designations, all of which are believed to indicate a substantial component of accreting envelope gas. PMS stars were considered to include "Class II" and "Class III" YSOs, where Class II YSOs include "transition disks." Class II YSOs are expected to have optically thick circumstellar disks with significant infrared excess above photospheric blackbody emission. Class III YSOs have relatively little infrared excess and are most easily detected by their x-ray emission. Relatively few regions have a complete census of Class III YSOs, so it was assumed that the number of Class III YSOs is equal to the number of Class IIs, following results of Luhman et al. (2008) in Cha I and Peterson et al. (2011) in Cr A. It will be important to complete the census of Class III YSOs in more embedded clusters, for more accurate results in the future.

The 23 clusters and eight complexes selected are listed in Table 2, in order of increasing number of protostars. Regions are considered clusters if they have one or a few neighboring local maxima of YSO surface density, and complexes if they have a more extended distribution of clusters. Complexes have more protostars than clusters, with 40 protostars separating the two types of region. In a few cases the classifications are arbitrary. The Lupus "cluster" is the sum of values in three Lupus clouds, after correction for extinction (Evans et al. 2009). The *c2d* "complex" is the sum of the *c2d* clouds Ophiuchus, Lupus, Cha II, Serpens, and Perseus (Evans et al. 2009). Where multiple studies of the same region are available, the more detailed study was generally selected, to avoid duplication.

The number of protostars and class II YSOs attributed to a cluster depends on the observed extent of the cluster, observational sensitivity, criteria for association, criteria for discriminating among classes, and on other factors. Therefore multiple studies of the same region find values of $N_{PS}$ or $N_{II}$ which may differ from one study to the next by a factor of $\sim 2$.

On the other hand, the ratio $N_{PS}/N_{II}$ varies by much less than a factor of 2 from one study to the next. Thus the cluster age, which depends on the ratio of protostars to YSOs, has smaller systematic error than the cluster birthrate, which is proportional to the total number of YSOs.

In this sample, the criteria for YSO association and classification also vary from one study to the next. It would be useful to apply the same criteria to all of the observations, but this improved procedure is beyond the scope of this paper.

The random error on cluster age is negligibly small compared to the systematic error. If the random uncertainty in $N_{PS}$ and in $N_{II}$ varies as $\sqrt{N}$ the fractional random error in $t$ is approximately equal to $N_{PS}^{-3/2}$. This fractional error is less than 3% when $N_{PS}$ exceeds 10.

The populations and model predictions are shown in Figure 9, a log-log plot of $N_{PS}$ vs. $N_{II}$. Figure 11 shows that $N_{PS}$ and $N_{II}$ are correlated when both clusters and complexes are considered. In terms of the constant birthrate model, the correlation means that the clusters and complexes have a much smaller range of ages than birthrates. The 23 clusters have mean ± standard deviation in their global age 2 ± 1 Myr, and in their birthrate 120 ± 60 protostars Myr$^{-1}$. The clusters and complexes taken together have the same mean and standard deviation in age as do the clusters alone. In contrast, the typical complex birthrate is ~ 1000 protostars Myr$^{-1}$, greater than the typical cluster birthrate by an order of magnitude.

The Serpens South cluster is remarkable in its high ratio of protostars to Class II YSOs, indicating a star-forming age less than ~ 0.5 Myr and a birthrate greater than 300 protostars Myr$^{-1}$. This region is also remarkable in the spatial structure of its protostar fraction, as discussed in Section 5.4.

**Table 2.** Ages and protostar birthrates of nearby clusters and complexes

| Cluster   | $N_{PS}$ | $N_{II}$ | ref | t (Myr) | b (YSOs Myr$^{-1}$) |
|-----------|----------|----------|-----|---------|---------------------|
| G81.37    | 4        | 35       | 1   | 3.1     | 24                  |
| G82.58    | 8        | 34       | 1   | 1.6     | 47                  |
| Cep A     | 8        | 46       | 2   | 2.1     | 47                  |
| G81.48    | 10       | 123      | 1   | 4.3     | 59                  |
| IC348     | 11       | 121      | 3   | 3.9     | 65                  |
| Lupus     | 12       | 54       | 4   | 1.7     | 71                  |
| G82.55    | 12       | 137      | 1   | 4.1     | 71                  |
| Cha I     | 14       | 94       | 5   | 2.4     | 83                  |
| GGD 12    | 15       | 63       | 2   | 1.6     | 89                  |
| G81.51    | 17       | 238      | 1   | 4.9     | 100                 |
| Cr A      | 19       | 43       | 6   | 0.92    | 110                 |
| L988e     | 19       | 73       | 7   | 1.5     | 110                 |
| LkHα 101  | 19       | 110      | 8   | 2.1     | 110                 |
| L1688     | 20       | 154      | 3   | 2.8     | 120                 |
| G82.57    | 22       | 30       | 1   | 0.60    | 140                 |
| G81.44    | 22       | 110      | 1   | 1.8     | 130                 |
| IC 5146   | 23       | 69       | 9   | 1.2     | 140                 |
| Mon R2    | 23       | 109      | 2   | 1.8     | 140                 |
| AFGL 490  | 28       | 90       | 2   | 1.2     | 170                 |
| I20050-1  | 29       | 55       | 2   | 0.79    | 180                 |
| Ser       | 30       | 198      | 10  | 2.4     | 180                 |
| NGC 1333  | 35       | 102      | 3   | 1.1     | 210                 |
| Ser S Complex | 37 | 11 | 11 | 0.18 | 330 |
| Taurus | 42 | 184 | 12 | 1.7 | 250 |
| Cep OB3 | 100 | 887 | 10 | 3.2 | 590 |
| RCW 38 | 113 | 437 | 13 | 1.5 | 670 |
| Carina | 164 | 514 | 14 | 1.2 | 980 |
| W 5 | 179 | 1890 | 15 | 3.8 | 1100 |
| Orion A | 229 | 2070 | 10 | 3.2 | 1300 |
| c2d | 251 | 635 | 4 | 1.0 | 1500 |
| Cyg | 870 | 7360 | 1 | 3.0 | 5100 |

_________________________________________

References - 1, Beerer et al. (2010); 2, Kryukova et al. (2012); 3, Jørgensen et al. (2009); 4, Evans et al. (2009); 5, Luhman et al. (2008); 6, Peterson et al. (2011); 7, Allen et al. (2008), 8, Wolk et al. (2010); 9, Harvey et al. (2008); 10, Kryukova et al. (2012); 11, Gutermuth et al. (2008); 12, Luhman et al. (2010); 13, Winston et al. (2011); 14, Smith et al. (2010); 15, Koenig et al. (2008).

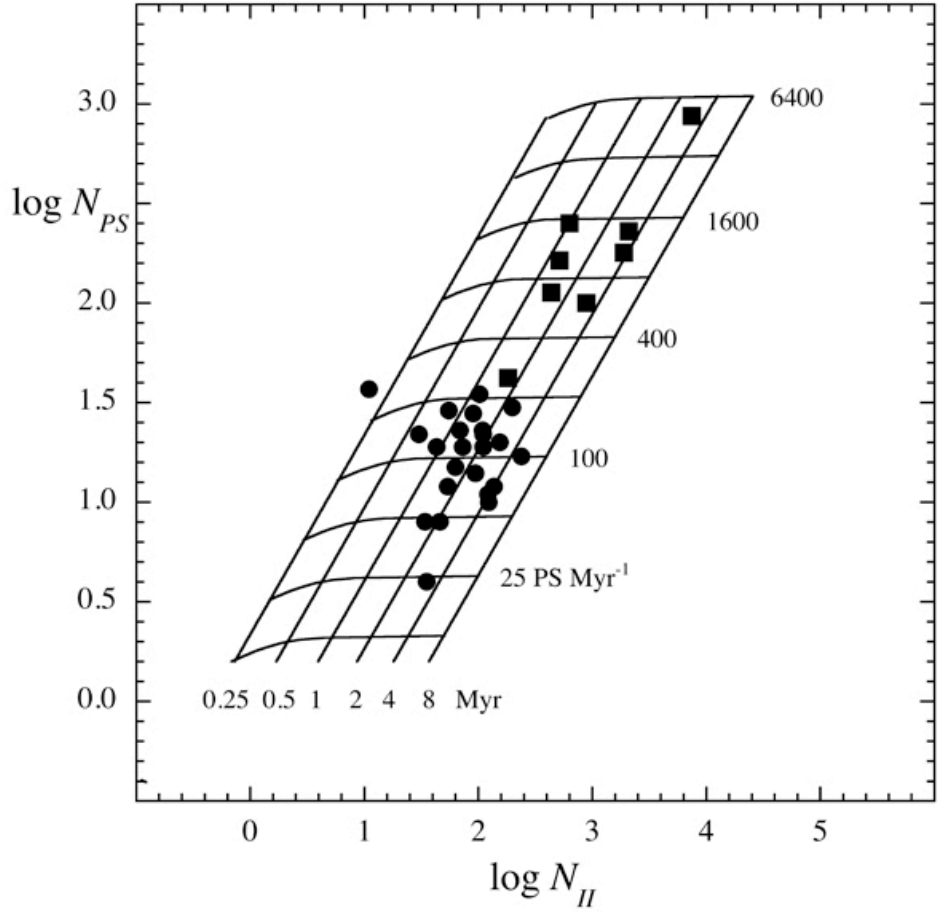

**Figure 9.** Comparison of the observed number of protostars $N_{PS}$ and the number of Class II YSOs $N_{II}$ in 23 nearby embedded clusters *(circles)*, and eight complexes *(squares)* with predictions of age and birthrate for the cluster model of constant birthrate, core-clump accretion, and equally likely stopping. It is assumed that the final YSO mass distribution matches the IMF and that the protostar luminosity distribution has modal value 1 $L_\odot$. This comparison indicates that most nearby clusters have star-forming age 1-3 Myr and birthrate 50-200 protostars $\text{Myr}^{-1}$. The complexes have similar ages but much greater birthrates.

The embedded cluster ages obtained here are comparable to young cluster ages obtained from optical spectroscopy and evolutionary models, having typical values of a few Myr, for example 1 Myr in L1688 in Ophiuchus (Luhman & Rieke 1999), 1.5 Myr in Taurus (Briceño et al. 2002), 2 Myr in Cha I (Luhman et al. 2008), and 3 Myr in IC348 (Luhman et al. 2006). These ages are generally mean or median ages and are uncertain by ~ 1 Myr. The fraction of YSOs in 14 clusters having optical evidence of accretion in the width of the Hα line declines with age, with an exponential time scale 2.3 Myr, and similarly the fraction of YSOs in 9 clusters having infrared evidence of an optically thick disk declines with an exponential time scale 3 Myr (Fedele et al. 2010). These estimates are sensitive to the PMS population in each region, but do not reflect the presence of protostars with accreting envelopes.

### 5.4. Age structure of the youngest clusters

Age estimation by the protostar fraction (PF), as described above, is especially useful for young clusters whose protostars are more concentrated than their PMS stars. In several such clusters, protostars dominate the YSO population in at least one small dense zone. Each such zone is surrounded by a more extended, less dense region dominated by PMS stars. This property is seen in IC 348-SW (Muench et al 2008), Serpens South (Gutermuth et al. 2008), Serpens (Gorlova et al. 2009), Corona Australis (Peterson et al. 2011), S140 N and L1211 (Gutermuth et al. 2009), and AFGL 490 Masiunas et al. (2012). For these and similar clusters, as the size of the zone considered increases, the local PF decreases and the local age increases, from a few 0.1 Myr in each protostar-dominated zone to ~ 1 Myr for the cluster as a whole.

For example, the Serpens North cluster and its YSOs presented by Gorlova et al. (2009) can be described as an enlongated zone of high extinction and 0.3 pc extent. It contains all 15 of the Serpens protostars and 15 of its PMS stars. This region is surrounded by less dense, filamentary gas extending over 0.9 pc, with 37 PMS stars. This region is surrounded by a more diffuse halo extending over 1.3 pc, with 31 PMS stars. In these three concentric zones, the protostar and PMS populations are reproduced by the PF model with mean accretion duration 0.16 Myr as in Section 5, and with birthrate 94 protostars $Myr^{-1}$, typical of the clusters in Table 2. The PF ages of the three zones increase outward from 0.3 to 0.7 to 1.0 Myr. In comparison, age estimates were made for 16 YSOs based on optical spectroscopy, yielding an average age ~ 1 Myr. Several of the YSOs closest to the cluster center lie above the 1 Myr isochrone of the HR diagram, i.e. they may be younger than 1 Myr. But it is difficult to assign specific ages to them because evolutionary tracks for ages less than 1 Myr are considered highly uncertain (Gorlova et al 209).

Similarly, the Serpens South cluster (Gutermuth et al 2008) and the CrA cluster (Peterson et al 2011) also have concentrations of protostars in a central zone with high surface density of gas and YSOs, surrounded by a filamentary structure with lower density gas and YSOs, surrounded by a halo of YSOs with little extinction. Applying the same procedure as above yields a progression of local ages 0.3 to 0.5 Myr in Serpens South and 0.4 to 0.6 to 0.9 Myr in CrA, each with birthrates near 100 $\text{Myr}^{-1}$.

These results suggest that in each region, some stars began forming with relatively low surface density as long as 1 Myr ago, while most of the stars in the densest part of the cluster began forming more recently. In the dense central zone, the inferred local age 0.3-0.4 Myr is consistent with the accretion time scale 0.2 Myr, since a high fraction of the YSOs are protostars, which must have been born only 1-2 accretion time scales ago.

This result might be consistent with a star formation history where the gas first became dense enough to make low-mass stars in a relatively extended filamentary configuration, while the more central region became dense enough to make clustered stars more recently. It will be useful to apply this method of PF age estimation to more clusters having well-defined protostar-dominated zones.

## 6. Discussion

This paper develops and applies the model of constant birthrate, core-clump accretion, and equally likely stopping first presented in Paper 1. The model is used here to predict the time evolution of accretion durations, population numbers, masses of PMS stars and protostars, and luminosities of protostars, during and after cluster formation. In contrast, Paper 1 treated only protostars during cluster formation. This work applies these models to fit the distribution of protostar luminosities in Orion A, yielding estimates of cluster age, birthrate, mean accretion duration, and core component of the mass accretion rate.

This paper further applies the model to estimate the cluster age and birthrate for 31 nearby embedded clusters and complexes where number of protostars and PMS stars are known, but the distribution of protostar luminosities is not known. The number of protostars is correlated with the number of PMS stars when both clusters and complexes are considered. The correlated populations have a relatively narrow range of ages 1-3 Myr, in good accord with estimates from optical spectroscopy.

This section discusses the limitations and implications of these models, and compares to related work. Additional related topics include isolated and clustered models of star formation, the relation of core masses to YSO masses, the basis of equally likely stopping, and the role of episodic accretion in protostar luminosities. These topics are discussed in Paper 1.

## 6.1. Limitations

The models of Sections 2-4 assume that variation in accretion duration from one protostar to the next is the most important factor in setting the resulting protostar mass distribution. With this assumption, all protostars are approximated here as having negligible variation in their accretion rate as a function of time, from one protostar to the next. However, the accretion rate will necessarily vary over the protostar population, due to observed differences in mean density and in density profiles from one core to the next. Such differences among the cores of a cluster-forming region can be seen in the profiles in the Lupus 3 cores (Teixeira et al. 2005). Also, the accretion rate can vary due to structure and evolution of the surrounding clump gas. The central condensation of clump gas provides a greater accretion rate close to its center of gravity than in more distant locations. Clump gas could provide a greater accretion rate at early times, if the clump is formed from supersonic colliding flows (e.g. Banerjee et al. 2009), or at later times, if the clump collapses due to global self-gravity (e.g. Burkert & Hartmann 2004).

It is difficult to include these more realistic features directly in an analytic model. However it would be useful to analyze cluster-forming simulations for the relative contributions of their accretion rate and of their accretion duration to the resulting distribution of protostar masses, extending analysis of this type in Bate (2011).

The birth history of protostars in a cluster is described here as production of protostars at a constant rate, with sudden onset and sudden stopping. This constant rate is set as a model parameter, and not as a physically determined quantity. Some studies of nearby regions indicate an accelerating rate of star formation (Palla & Stahler 2000). Such acceleration has been attributed to gravitational contraction of a cluster-forming clump (Huff & Stahler 2006, 2007), and has been described analytically by OM11. The approach to a constant population of protostars found here and in FS94 is a feature of the constant birthrate model combined with equally likely accretion stopping. In contrast, for an exponentially increasing birthrate, the population of protostars is not constant. Instead the number of protostars increases monotonically with time, while the protostar fraction approaches a fixed ratio which depends on the time scales of birthrate increase and of accretion stopping.

The present model is simplified because it describes a single episode of star formation, whereas observed embedded clusters are rarely isolated in space and time. Instead, many clusters belong to complexes, as in Orion A, Cygnus, and Carina. Studies of these complexes indicate multiple episodes of star formation within a cluster, and multiple episodes of cluster formation within a complex, spread out over tens of pc and over time spans exceeding ~ 10 Myr (Smith et al. 2010, De Marchi et al. 2011).

The age estimates in Section 5 are limited by incompleteness in knowledge of the populations of protostars and PMS stars associated with a cluster or complex. The YSO population of a cluster is uncertain because clusters do not have a clear defining boundary (Bressert et al. 2010). Even if observational factors were negligible, the number of associated YSOs would vary depending on how the outermost members were defined. In addition, confusion with background sources makes determination of membership uncertain, especially at low flux levels (e.g. Harvey et al. 2007). These factors increase the relative uncertainty in the protostar birthrate of an embedded cluster in proportion to the relative uncertainty of the population. These factors have less effect on the age estimate, which depends more on the ratio of protostar and PMS populations.

Determination of whether an associated YSO is an accreting protostar, or a non-accreting PMS star is also uncertain because classification is generally determined by infrared colors (Allen et al. 2007) and not by a measurement of accretion rate. More detailed fitting to spectral energy distributions appears more accurate than placement on color-color diagrams, but still generates a significant fraction of ambiguous classifications (Smith et al. 2010). Furthermore, current knowlege of Class III source populations in embedded clusters is limited to a relatively small number of clusters whose x-ray emission has been studied in conjunction with their infrared emission (Winston et al. 2010, 2011, Wolk et al. 2010). Assignment of equal numbers of Class II and Class III YSOs, as was assumed in Section 5, is justified by some but not all studies of Class III sources, with younger clusters tending to have relatively fewer Class III members and older clusters having relatively more Class III members. These uncertainties in relative populations of protostars and PMS stars suggest a factor of $\sim 2$ uncertainty in the estimated age of an embedded cluster.

## 6.2. Implications

Despite the uncertainties described above, the model of constant birthrate, core-clump accretion, and equally likely stopping presented here provides a simple way to describe protostars and PMS stars during cluster formation. It gives a good fit to the IMF and to the luminosity distribution of protostars in Orion A. This fit yields an estimate of star-forming age, 3 Myr, comparable to that of the ONC, and yields estimates of mass accretion rate and mean accretion duration in good agreement with independent determinations for regions of low-mass star formation.

The model provides a simple interpretation to the correlation of protostar and Class II populations in embedded clusters and complexes, in terms of a narrow range of star-forming ages, 1-3 Myr. These ages based on observed protostar fractions are similar to typical PMS star ages in young clusters, determined from optical spectroscopy of YSOs and stellar evolution models. It will be useful to make more detailed comparisons of cluster ages according to these two methods.

It is notable that embedded clusters with a global protostar fraction exceeding 0.5 are rare according to the literature survey summarized in Figure 9. This property may arise because a cluster must have an unusually high birthrate in order to have similar numbers of protostars and PMS stars, and to have enough YSOs to be considered a cluster. Equations (7) and (8) indicate that a cluster of at least 35 observed members with equal numbers of protostars and PMS stars must have a birthrate at least ~170 protostars Myr$^{-1}$. This birthrate is greater than nearly all of the cluster birthrates in Figure 9. Indeed the Serpens South cluster, with ~300 protostars Myr$^{-1}$, is both the youngest and the highest-birthrate cluster in the sample.

The spatial structure of the protostar fraction in very young clusters, which cannot be obtained from OS studies, may constrain the history of forming clusters. Section 5.4 shows that in some young clusters, the region which is densest in gas and stars also has the greatest protostar fraction. This combination suggests that most star formation in such localized zones started more recently and has proceeded more efficiently than in the surrounding gas. The inferred star-forming age of a few 0.1 Myr is comparable to the free-fall time of cluster-forming gas with a mean density $10^4$ cm$^{-3}$. If the onset of such star formation is due mainly to the availability of sufficiently dense gas, the high protostar fraction may in turn indicate that its parent gas became sufficiently dense more recently than the birth time of the oldest stars in the surrounding complex.

## 7. Conclusion

This paper presents an analytic description of the birth history, masses and luminosities of YSOs in clusters. In a forming cluster, a parsec-scale clump harbors closely spaced cores and accreting protostars. The protostar mass accretion rate has a constant "core" component and a mass-dependent "clump" component, which together account for formation of low-mass and massive stars. The final mass of a protostar depends primarily on processes which limit accretion, including dynamical ejection of protostars from small multiple systems, competition with nearby accretors, and the ionization, heating, and outflows due to nearby young stars. The distribution of accretion durations is described by the stochastic model of equally likely stopping.

These ideas were developed in Paper 1 and in related papers, including Palla & Stahler (2000) for the time distribution of protostar births, Myers & Fuller (1992) and McKee & Tan (2003) for core-clump accretion, and Basu & Jones (2004) and Bate & Bonnell (2005) for equally likely stopping. Paper 1 predicted distributions of protostar masses and accretion luminosities as a function of cluster age in a constant-birthrate cluster, as the protostar population approaches a steady state, similar to that of FS94.

This paper extends the results of Paper 1 to give a more detailed picture of YSO evolution in clusters. It presents time-dependent distributions of accretion age for protostars and PMS stars, during and after cluster formation. These properties of accretion duration are used to predict distributions of YSO mass and luminosity and to estimate cluster ages. The main results are:

1. The number of protostars approaches a steady state after a few accretion time scales, as the rate of accretion stopping approaches the birth rate. In contrast the number of PMS stars increases monotonically. This differing time dependence allows estimates of the star-forming age of a cluster.

2. The distributions of protostar and PMS mass, and of protostar luminosity, evolve as the cluster develops. Each distribution approaches a steady state having a steep rise, a peak, and a shallow decline. The maximum mass and luminosity each increase with time.

3. The distribution of YSO masses after cluster formation matches the IMFs of Kroupa (2002) and Chabrier (2005) provided the accretion model parameters have the adopted values $m_0 = 0.34\ M_\odot$, $q = 2.0$, and $p = 1.2$.

4. During cluster formation, the steady-state protostar mass distribution has the same shape as the YSO final mass distribution after cluster formation.

5. The distributions of protostar mass $dN/d\log m$ and accretion luminosity $dN/d\log L$ have nearly identical dependence on mass. The modal mass and the modal luminosity are simply expressed in terms of the accretion time scale and the core mass accretion rate.

6. The predicted distribution of protostar luminosities matches the distribution in the Orion A cloud reported by Kryukova et al (2012). The fit parameters yield star-forming age 3 Myr and the protostar birthrate 1500 protostars $Myr^{-1}$. This Orion A age is similar within errors to the age of the ONC estimated from optical spectroscopy and evolutionary tracks (Reggiani et al. 2011).

7. The star-forming age and birthrate are estimated for 31 embedded clusters and complexes, based on their numbers of protostars and Class II YSOS, and assuming the same modal luminosity as in Orion and other well-studied complexes. The typical global ages are 1-3 Myr, similar to those based on optical spectroscopy. The typical birthrates are 60-180 protostars Myr$^{-1}$ in clusters and ~1000 protostars Myr$^{-1}$ in complexes.

8. The youngest, most obscured clusters have dense central zones whose protostar fraction is a local maximum. These zones are surrounded by less dense, extended zones dominated by PMS stars. The protostar fraction gives the best available way to estimate ages in these clusters, since optical spectroscopy is not sensitive to the protostars. In Serpens South, Serpens North, and Corona Australis, the inferred star-forming age increases from a few 0.1 Myr in the central zone to ~ 1 Myr when all YSOs are considered. This result suggests that the central parts of these regions have become dense enough to form clustered protostars only in the last few 0.1 Myr.

**Acknowledgements**  The author thanks Fred Adams, Lori Allen, João Alves, Shantanu Basu, Matthew Bate, Rob Gutermuth, Helen Kirk, Erin Kryukova, Charlie Lada, Chris McKee, Stella Offner, Ralph Pudritz, Howard Smith,  Loredana Spezzi, and Qizhou Zhang for helpful discussions.  Terry Marshall and Irwin Shapiro provided support and encouragement.  The referee provided helpful comments which improved the focus of the paper.